\documentstyle[12pt,epsf]{article}
\oddsidemargin   - 1.1cm
\begin{document}
\pagestyle{empty}

\begin{center}
SIGN REVERSAL OF THE QUANTUM HALL EFFECT AND HELICOIDAL
MAGNETIC-FIELD-INDUCED SPIN-DENSITY WAVES IN ORGANIC CONDUCTORS
\end{center}

\centerline{N. Dupuis$^{(1)}$ and Victor M. Yakovenko$^{(2)}$}

\begin{center}
$^{(1)}$Laboratoire de Physique des Solides, Bat. 510, 
Centre Universitaire, 91405 Orsay C\'edex, France  \\
$^{(2)}$Department of Physics, University of Maryland, College Park
20742-4111, USA 
\end{center}

\begin{center}
{\it Proceedings of the International Workshop on electronic
crystals (ECRYS99)}  
\end{center} 

\vspace{1cm}

\noindent
\hfill  \parbox{16.5cm}{
\footnotesize {\bf Abstract.}
Within the framework of the quantized nesting model, we study the
effect of umklapp scattering on the magnetic-field-induced 
spin-density-wave (SDW) phases which are experimentally observed in the
quasi-one-dimensional organic conductors (TMTSF)$_2$X. We discuss the
conditions under which umklapp processes may explain the sign
reversals (Ribault anomaly) of the quantum Hall effect (QHE)
observed in these conductors. We find that the `Ribault phase' is
characterized by the coexistence of two SDWs with comparable
amplitudes. This gives rise to additional long wavelength collective
modes besides the Goldstone modes due to spontaneous translation and
rotation symmetry breaking. These modes strongly affect the optical
conductivity. We also show that the Ribault phase may become
helicoidal (i.e with circularly polarized SDWs) if the strength of
umklapp processes is sufficiently strong. The QHE vanishes in the
helicoidal phases, but a magnetoelectric effect appears.
}

\vspace{.9cm}

\noindent
{\bf 1. INTRODUCTION}
\vspace{.5cm}

\noindent
The organic conductors of the Bechgaard salts family (TMTSF)$_2$X (where
TMTSF stands for tetramethyltetraselenafulvalene) have remarkable properties
in a magnetic field. In three members of this family (X=ClO$_4$, PF$_6$,
ReO$_4$), a moderate magnetic field of a few Tesla destroys the
metallic phase and induces a series of SDW phases
separated by first-order phase transitions [1].

According to the so-called quantized nesting model (QNM) [1],
the formation of the FISDWs results from the
strong anisotropy of these organic materials, which can be viewed as weakly
coupled chain systems (the typical ratio of the electron
transfer integrals in the three crystal directions is
$t_a:t_b:t_c=3000:300:10$ K). The SDW opens a gap, but leaves
closed pockets of electrons and/or holes in the vicinity of the Fermi
surface. In presence of a magnetic field $H$, these pockets are
quantized into Landau levels (more precisely Landau subbands). In each FISDW
phase, the SDW wave vector is quantized, ${\bf Q}_N=(2k_F+NG,Q_y)$ with $N$
integer, so that an integer number of Landau subbands are filled. [Here $k_F$
is the Fermi momentum along the chains, $-e$ the electron charge, $b$ the
interchain spacing, and $G=eHb/\hbar $.] As a result, the Fermi level lies
in a gap between two Landau subbands, the SDW phase is stable, and the Hall
conductivity is quantized: $\sigma _{xy}=-2Ne^2/h$ per one layer of the
TMTSF molecules [1]. As the magnetic field
increases, the value of the integer $N$ changes, which leads to a cascade of
FISDW transitions. 

A striking feature of the QHE in Bechgaard salts is the coexistence of both
positive and negative Hall plateaus. While most plateaus are of the same
sign, referred to as positive by convention, a negative Hall effect is also
observed at certain pressures (the so-called Ribault anomaly) [2].  

We have recently proposed an explanation of the Ribault anomaly by taking
umklapp processes into account [3,4]. The latter are allowed in the
Bechgaard salts due to the half-filled electron band which results
from the dimerization along the chains. Within our explanation [5], two
SDWs with comparable amplitudes  coexist in the negative phases, which
therefore differ significantly from the positive ones. In particular,
they exhibit an unusual structure of long-wavelength collective modes [4],
and their polarization may become circular (helicoidal SDWs) above a
critical value of the umklapp scattering strength [3]. 
\vspace{.6cm}

\newpage

\noindent
{\bf 2. SIGN REVERSAL OF THE QHE} 
\vspace{.5cm}

\noindent
In the absence of umklapp scattering, a magnetic field along the $z$
axis induces a series of SDW phases characterized by a quantized wave
vector ${\bf Q}_N=(2k_F+NG,Q_y)$ and a quantification of the Hall
effect: $\sigma_{xy}=-2Ne^2/h$. The sign of $N$ is entirely determined
by the electron dispersion [1], which seems to preclude any sign
reversal of the QHE. 

Umklapp scattering transfers $4k_F$ and therefore couples the wave
vectors ${\bf Q}_N$ and ${\bf Q}_N-4k_F$. As a result two SDWs, with
wave vectors ${\bf Q}_N=(2k_F+NG,Q_y)$ and ${\bf Q}_{-N}=(2k_F-NG,-Q_y)$
form simultaneously. In the random-phase approximation, the transition
temperature is determined by the modified Stoner criterion 
\begin{equation}
[1-g_2\chi_0({\bf Q}_N)] [1-g_2\chi_0({\bf Q}_{-N})]
-g_3^2  \chi_0({\bf Q}_N)\chi_0({\bf Q}_{-N})= 0,
\end{equation}
where $g_2$ and $g_3$ are the scattering amplitudes of the normal and
umklapp processes, respectively. $\chi_0$ is the spin
susceptibility in the absence of electron-electron interaction. 

A very small $g_3$ does not qualitatively change the phase diagram
compared to the case $g_3=0$. Now the main SDW at wave vector
${\bf Q}_N$ coexists with a weak SDW at wave vector ${\bf
  Q}_{-N}$. The values of $N$ follow the usual `positive'
sequence $N=\ldots,4,3,2,1,0$ with increasing magnetic field.
A larger value of $g_3$ increases the coupling between the two SDWs. This
leads to a strong decrease of the transition temperature or even the 
disappearance of the SDWs. However, for even $N$, there exists a
critical value of $g_3$ above which the system prefers to choose the
transversely commensurate wave vector $Q_y=\pi/b$ for both SDWs. [This
follows from the structure of the bare spin susceptibility $\chi _0({\bf
Q})$ at $Q_y=\pi /b$.] The two SDWs have now comparable
amplitudes. For certain dispersion laws [6], the 
main SDW corresponds to the wave vector ${\bf
Q}_{-|N|}$, which yields a negative Hall plateau. Thus, for $
g_3/g_2\simeq 0.03$, we find the sequence
$N=\dots,5,4,-2,2,1,0$ (Fig.\ 1a). In Bechgaard salts, the
strength of umklapp scattering is very sensitive to pressure. Therefore, we
conclude that sign reversals of the QHE can be induced by varying pressure
as observed experimentally [2]. 

\begin{figure}[h]
\epsfysize 5.8cm 
\epsffile[50 600 395 750]{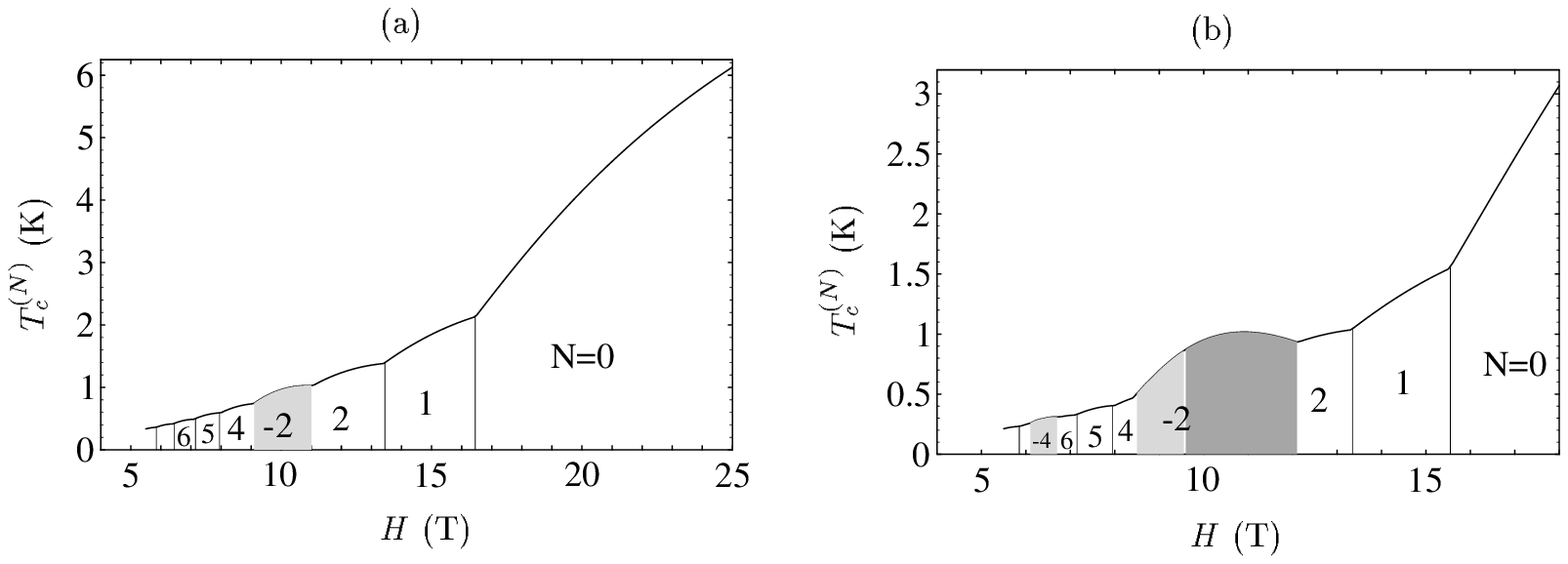}
\label{Fig}
\end{figure}

\noindent
{\footnotesize {\bf Figure 1:} (a) Phase diagram for $g_3/g_2=0.03$.
  The phase $N=3$ is suppressed, and the
  negative commensurate phase with $N=-2$ and $Q_y=\pi/b$ appears in
  the cascade (the shaded area). All the phases are sinusoidal, and
  the Hall effect is quantized: $\sigma_{xy}=-2Ne^2/h$. The vertical
  lines are only guides for the eyes and do not necessarily correspond to
  the actual first-order transition lines. 
 (b) Phase diagram for $g_3/g_2=0.06$. Two negative phases, $N=-2$ and
  $N=-4$, are observed (the shaded areas). The phase $N=-2$ splits
  into two subphases: helicoidal (the dark shaded area) and sinusoidal
  (the light shaded area). }
\vspace{.6cm}

\noindent
{\bf 3. HELICOIDAL PHASES}
\vspace{.5cm}

We have also shown that umklapp scattering can change the polarization of
the SDWs from linear (sinusoidal SDWs) to circular (helicoidal SDWs)
[3]. The QHE vanishes in the helicoidal phases, but a magnetoelectric
effect appears. For  $g_3/g_2\simeq 0.06$, 
there are two negative phases: $N=-2$ and $N=-4$ (Fig.\ 1b). The phase 
$N=-2$ splits into two subphases: helicoidal and
sinusoidal. In order to observe helicoidal phases experimentally, it would
be desirable to stabilize the negative phase $N=-2$ at the lowest possible
pressure (which corresponds to the strongest $g_3$). 
\vspace{.6cm}

\noindent
{\bf 4. COLLECTIVE MODES}
\vspace{.5cm}

The Ribault phase exhibits an unusual structure of long-wavelength
collective modes. The coexistence of two SDWs in this phase gives rise
to additional modes besides the Goldstone modes resulting from
spontaneous rotation and translation symmetry breaking [4]. Below we
discuss only the sliding modes, restricting ourselves to longitudinal
(i.e. parallel to the chains) fluctuations. 

There is a Goldstone mode with a linear dispersion law $\omega
=v_Fq_x$, which corresponds to out-of-phase oscillations of the two
SDWs: $\theta_N(q_x,\omega)=-\theta_{-N}(q_x,\omega)$, where the
phases $\theta_N$ and $\theta_{-N}$ determine the positions of the two
SDWs with respect to the crystal lattice. The fact that the two SDWs can be
displaced in opposite directions without changing the energy of the
system is related to the pinning that would occur for a single
commensurate SDW. 

Besides the Goldstone mode, there is a gapped mode $\omega^2=
\omega_0^2 + v_F^2q_x^2$, which corresponds to in-phase oscillations
of the two SDWs. $\omega_0$ depends on the strength of umklapp
scattering and is generally larger than the mean-field gap. We
therefore expect this mode to be strongly damped due to the coupling
with the quasi-particle excitations. 

The presence of two SDWs can be detected by measuring the optical
conductivity. In the limit ${\bf q}=0$, the dissipative part of the
conductivity is given by
\begin{equation}
{\rm Re}[\sigma(\omega)]=\frac{\omega_p^2}{4}\Biggl ( \delta(\omega)
\frac{3(1-\tilde\gamma^2)}{3+5\tilde\gamma^2}+\delta(\omega\pm\omega_0)
\frac{4\tilde\gamma^2}{3+5\tilde\gamma^2} \Biggr ),
\label{cond}
\end{equation}
where $\tilde\gamma$ is the ratio of the SDW amplitudes, and
$\omega_p$ the plasma frequency. Eq.~(\ref{cond}) satisfies the
conductivity sum rule $\int _{-\infty}^\infty {\rm Re}[\sigma
(\omega)]=\omega_p^2/4$. 
Quasi-particle excitations above the mean-field gap do not
contribute to the optical conductivity, a result well known in clean 
SDW systems. Because both modes contribute to the 
conductivity, the low-energy (Goldstone) mode carries only a fraction of the
total spectral weight. We obtain Dirac
peaks at $\pm \omega _0$ because we have neglected the coupling of the gapped
mode with quasi-particle excitations. Also, in a real system
(with impurities), the Goldstone mode would broaden and appear at a finite
frequency (below the quasi-particle excitation gap) due to pinning by
impurities.
\vspace{0.6cm}

\noindent
{\bf References}
\vspace{0.5cm}

\noindent
1. For recent reviews, see P. M. Chaikin, J. Phys. (Paris)
  I {\bf 6}, 1875 (1996); P. Lederer, {\it ibid}, p. 1899; V.  M.
  Yakovenko and H. S. Goan, {\it ibid}, p. 1917. \\
2. M. Ribault, Mol. Cryst. Liq. Cryst. {\bf 119}, 91
  (1985); L. Balicas  {\it et al.}, Phys. Rev. Lett. {\bf 75}, 2000 (1995).\\
3. N. Dupuis and V. M. Yakovenko, Phys. Rev. Lett. {\bf 80}, 3618
    (1998); Phys. Rev. B {\bf 58}, 8773 (1998). \\
4. N. Dupuis and V. M. Yakovenko, Europhys. Lett. {\bf 45}, 361 (1999). \\
5. Another explanation of the negative Hall plateaus, based on the
dependence of the electron dispersion on pressure, has been proposed
by D. Zanchi and G. Montambaux, Phys. Rev. Lett. {\bf 77}, 366
(1996). \\
6. The exact condition is $t_{2b}t_{4b}>0$, where $t_{2b}$ and
$t_{4b}$ are the second and fourth harmonics of the transverse
dispersion law. 

\end{document}